\documentclass[letter]{aa} % for the letters 
\usepackage{graphicx}
\usepackage{xcolor}
\usepackage{float}
\usepackage{chngcntr}
\usepackage{textcomp}
\usepackage{booktabs}

\usepackage[colorlinks]{hyperref}  
\hypersetup{citecolor=blue} 

\usepackage{txfonts}
\begin{document}

   \title{Toward 3D orbits of wide sdO/B binaries}

   \subtitle{I. Six composite systems spatially resolved with VLTI/GRAVITY}

    \author{K. Deshmukh \inst{\ref{inst:kul},\ref{inst:lgi}}\thanks{Corresponding author; \texttt{astro.kunal.deshmukh@gmail.com}} \and
           S. Geier \inst{\ref{inst:potsdam}}
           \and
           H. Sana \inst{\ref{inst:kul},\ref{inst:lgi}}
           \and 
           H. Dawson \inst{\ref{inst:potsdam}}
           \and
           M. Dorsch \inst{\ref{inst:potsdam}}
           \and
           A. J. Frost \inst{\ref{inst:eso-c}}
           \and 
           U. Heber \inst{\ref{inst:fau}}
           \and 
           T. Kupfer \inst{\ref{inst:ham},\ref{inst:ttu}}
           \and \\
           A. Picco \inst{\ref{inst:kul},\ref{inst:lgi}}
           \and
           M. Vučković \inst{\ref{inst:val}}
           }

   \institute{
   {Institute of Astronomy, KU Leuven, Celestijnenlaan 200D, 3001 Leuven, Belgium \label{inst:kul}}\\ 
   \email{astro.kunal.deshmukh@gmail.com}
   \and %LeuveGravityInstitute
    {Leuven Gravity Institute, KU Leuven, Celestijnenlaan 200D, box 2415, 3001 Leuven, Belgium \label{inst:lgi}}
    \and %Potsdam
    {Institute for Physics and Astronomy, University of Potsdam, KarlLiebknecht-Str. 24/25, 14476 Potsdam, Germany \label{inst:potsdam}}
    \and %Abi
    {European Southern Observatory, Alonso de Cordova 3107, Vitacura, Santiago, Chile \label{inst:eso-c}}
    \and %Uli
    {Dr. Remeis-Sternwarte and ECAP, Astronomical Institute, University of Erlangen-Nürnberg, Sternwartstr. 7, D-96049 Bamberg, Germany\label{inst:fau}}
    \and %Thomas
    {Hamburger Sternwarte, University of Hamburg, Gojenbergsweg 112, 21029 Hamburg, Germany \label{inst:ham}}
    \and 
    {Department of Physics \& Astronomy, Texas Tech University, Box 41051, Lubbock, TX, 79409-1051, USA \label{inst:ttu}}
    \and %Maja
    {Instituto de Física y Astronomía, Universidad de Valparaíso, Gran Bretaña 1111, Playa Ancha, Valparaíso 2360102, Chile\label{inst:val}}
    }

   \date{Received; Accepted}

% \abstract{}{}{}{}{} 
% 5 {} token are mandatory
 
  \abstract
  {Hot subdwarf stars (sdO/Bs) are widely considered to be products of binary evolution. A significant fraction of them are found in long-period or wide binaries ($P>500$\,d) with main sequence (MS) companions, likely resulting from a stable mass transfer episode where the MS companion stripped the hydrogen envelope of the sdO/B progenitor. Consequently, wide sdO/B binaries represent a key population in our pursuit of understanding stable mass transfer. They exhibit a modest range of orbital periods and eccentricities as revealed by long-term spectroscopic campaigns, though the component masses are not well constrained. In this Letter, we present the first long-baseline interferometry campaign to observe wide sdO/B + MS binaries and take the first step toward determining their 3-dimensional (3D) orbits and model-independent component masses. We target six composite sdO/B + MS systems with VLTI/GRAVITY and spatially resolve all of them. The projected physical separations range between $1-3$ au, with uncertainties between $1-8$\%. When combined with complementary information from spectroscopic or astrometric observations, our precise measurements will be crucial to constrain 3D orbits for these systems.  
  Additionally, we also identify a potential third component in BD\,$+10^\circ2357$, although additional data will be necessary for confirmation.
  In light of continued spectroscopic monitoring and the imminent \textit{Gaia} Data Release 4, we strongly encourage expanding the interferometric sample presented here to establish new, precise orbital and mass constraints for this key population of binary interaction products.}

   \keywords{stars: subdwarfs -- (stars:) binaries: general -- techniques: interferometric}

   \titlerunning{Six composite sdO/B binaries spatially resolved with VLTI/GRAVITY}
   
   \maketitle
%
%-------------------------------------------------------------------

\section{Introduction}

Hot subdwarf stars are stripped remnants of low to intermediate ($M_{\rm ini}\lesssim8\,M_\odot$) mass stars. They are understood to be helium-burning objects with a thin hydrogen envelope, giving them a spectral appearance similar to O- and B-type stars. As a result, they are classified as sdB or sdO, with further subclasses based on surface helium abundances \citep{2024Heber}. At the center of hot subdwarf (“sdO/B” here onward) formation is the stripping of most of their hydrogen envelope, which is thought to be possible only with the help of a companion \citep{2020Pelisoli}. According to binary population synthesis studies, as their progenitors expand to become red giants, they initiate a mass transfer episode with the companion which may evolve in a stable or unstable manner and produce a broad range of orbital periods \citep[$P\sim10^{-2}-10^3$\,d,][]{2002Han,2003Han,2020Vos}.

The spectral classification sdO/B, however, applies to a variety of objects with overlapping but different mass ranges, such as very short-lived post-asymptotic giant branch stars ($\sim$~$0.6-0.8\,M_{\rm \odot}$), (post-) core helium burning stars originating from low-mass or intermediate-mass progenitors ($\sim$~$0.47\,M_{\rm \odot}$ or $\sim$~$0.3-0.9\,M_{\rm \odot}$), or (pre-) helium white dwarfs without fusion in their cores ($\sim$~$0.2-0.4\,M_{\rm \odot}$). In contrast to O- or B-type main-sequence (MS) stars, the sdO/B spectral types or temperatures are not directly correlated to their masses. The current methods to determine sdO/B masses, however, which are dynamical measurements in single-lined, eclipsing or wide binaries \citep[e.g.][]{2021Schaffenroth,2026Molina}, asteroseismology \citep{2012Fontaine}, or combining spectroscopy, photometry, and {\it Gaia} parallaxes \citep[e.g.][]{2026Dawson}, are all to some extent model dependent and limited in accuracy. Accurate masses are essential in determining the evolutionary status of sdO/Bs and constraining their formation channels.

Long-period or wide sdO/B binaries represent an important sub-class that likely results from the stable mass transfer channel. The sdO/B stars in wide binaries typically have MS companions of spectral types ranging from A to K with orbital periods between 500 and 2000 days \citep{2026Molina}. They also display a modest distribution of eccentricities ($e \approx 0-0.2$) and pair-wise correlation between orbital period, mass ratio, and eccentricity \citep{2017Vos,2019Vos,2026Molina}. Given they likely form via stable mass transfer, constraining their orbital and stellar properties is crucial in informing mass transfer physics. 

Wide sdO/B binaries are often identified and characterized based on their photometric colors, spectral energy distribution (SED) analysis,
and spectroscopic analysis using spectral lines of both components \citep[e.g.][]{2003Stark,2018Heber}. As a result, they are also referred to as composite sdO/B binaries. 
%In many cases, the optical flux for sdO/B + MS binaries can be dominated by the MS component (early F- or A-type MS star) or by the sdO/B component (K-type MS star), making it difficult to obtain precise stellar parameters for both components.
In many cases, the optical flux from sdO/B + MS binaries can be dominated by the MS component when the companion is an early F- or A-type star, or by the sdO/B component when the companion is a K-type star. This can make it difficult to obtain precise stellar parameters for both components.
Furthermore, the low radial velocity (RV) amplitude accompanying their long periods makes constraining their orbital solutions difficult and time-consuming. Several wide sdO/B + MS binaries have been studied over the past decades, most commonly via dedicated RV follow-up studies \citep[e.g.][]{1990Howarth,2012Deca,2017Vos,2026Molina}. While the orbital periods and RVs of the MS component are determined precisely, the sdO/B RVs are harder to measure due to fewer and broader spectral lines compared to its companion, as well as the absence of strong metal lines. As a result, even if the MS component mass can be estimated accurately based on spectroscopic modeling, the subsequent sdO/B mass estimate is rendered uncertain by an imprecise mass ratio. 

The {\it Gaia} space mission can alleviate some of the challenges of spectroscopy by providing complementary information. In {\it Gaia} Data Release 3 \citep[DR3,][]{2023aGaia,2023bGaia} a handful of wide sdO/B + MS binaries were identified using astrometry as part of its non-single star catalog when cross-matched with the samples of \citet{2019Geier} and \citet{2022Culpan} \citep[see also][]{2026Molina}. Many more are expected to be identified in the fourth data release (DR4) later this year \citep{2026Nagarajan}. While these binaries are not spatially resolved in {\it Gaia} observations, their photocenter motion around the binary center-of-mass can provide all orbital elements except the semi-major axis. As a result, a full determination of the 3-dimensional (3D) orbit is contingent upon estimating an accurate flux ratio between the two components.

Alternatively, directly measuring the relative positions of the binary components can also constrain the full orbit when combined with an astrometric orbital solution from {\it Gaia}. Long-baseline interferometry is a powerful and efficient technique to make such a measurement and lead us to model-independent dynamical masses of the components. Multi-epoch interferometric observations with sufficient orbital phase coverage can enable even more precise orbital constraints. This method has been used successfully for Be stars with subdwarf companions \citep{2024Klement,2025Klement}, also motivating focused theoretical studies using such systems to understand mass transfer physics \citep{2025Lechien,2026Picco}. As a result, long-baseline interferometry can open a new window to study sdO/B + MS binaries.

In this Letter, we present the first long-baseline interferometric observations of six sdO/B stars in wide binaries with MS companions. We spatially resolve all six binaries and demonstrate the effectiveness of interferometry to study these post-interaction systems. The Letter is structured as follows: the observations and data reduction are discussed in Section\,\ref{sec:obs}, followed by the analysis of interferometric data to resolve the binaries described in Section\,\ref{sec:comp}. We conclude with Section\,\ref{sec:dis}, where we discuss future prospects for interferometric observations of wide sdO/B binaries and synergy with complementary observational methods.

\section{Observations and data reduction}
\label{sec:obs}

The GRAVITY instrument \citep{2017GRAVITY} at the Very Large Telescope Interferometer (VLTI) is a powerful tool to study binaries. Operating in the $K$-band region (1.97--2.45 $\mu$m), it is sensitive to binary separations between $\sim$~$1-100$ milliarcseconds (mas) and flux contrasts up to $\Delta K\approx 5$\,mag. We observed a sample of six composite sdO/B binaries using the single-field on-axis mode of VLTI/GRAVITY between October 2025 and May 2026. Two of the targets, HD~128220 and BD\,$-07^\circ5977$, have orbital information available from the literature \citep{1990Howarth,2017Vos}. The remaining four were classified as composite binaries based on SED analysis \citep{2022Culpan,2024Dawson}. Three targets with Vega magnitudes $K<9$ were observed with the array of four 1.8-m Auxiliary Telescopes (ATs), while the remaining three with $K>9$ were observed with the four 8.2-m Unit Telescopes (UTs). The configuration A0-G1-J2-K0 was used for AT observations, with the longest baseline extending to around 130 m.

Calibrator stars were selected using the SearchCal tool \citep{2011Bonneau}, based on close proximity to the science target, similar $K$-band magnitude, and a K-type giant spectral class. A summary of the observations is given in Appendix\,\ref{app:obs}. All observations were processed and calibrated with the GRAVITY data reduction pipeline \citep[version 1.9.6;][]{2014Lapeyrere} using the ESO Data Processing System \citep{2024Freudling}.

\section{Data analysis and results}
\label{sec:comp}

We retrieved the visibility amplitudes (|V|) and closure phases (T3PHI) for all targets to perform the binary companion search analysis. |V| and T3PHI encode information about the spatial intensity distribution of the target, and are thus powerful diagnostic tools to detect binaries or higher order systems. In simple terms, |V| captures the extent of how spatially resolved the source is and the flux contrast between different components (if multiple components present), while T3PHI is sensitive to asymmetries in the spatial intensity distribution. For resolved binaries or multiples, |V| and T3PHI can enable us to constrain the angular separation, orientation, and flux ratio between the components.

   \begin{table*}
      \caption[]{Results from companion search for GRAVITY observations of six composite sdO/B binaries. }
         \label{tab:results}
         \renewcommand{\arraystretch}{1.3}
\begin{tabular*}{\textwidth}{@{\extracolsep{\fill}} ccccccccc }
\hline
\begin{tabular}{@{}c@{}} Target  \end{tabular} 
& \begin{tabular}{@{}c@{}} $d$ [pc] \end{tabular}  
& \begin{tabular}{@{}c@{}} ($\Delta E,\Delta N$) [mas] \end{tabular} 
& \begin{tabular}{@{}c@{}} ($e_{\rm maj},e_{\rm min}$) [mas] \end{tabular} 
& \begin{tabular}{@{}c@{}} PA [deg] \end{tabular} 
& \begin{tabular}{@{}c@{}} $\chi^2_{\rm red}$ \end{tabular}
& \begin{tabular}{@{}c@{}} $\rho$ [au] \end{tabular}
& \begin{tabular}{@{}c@{}} $f_2/f_1$ [\%] \\ \end{tabular}
& \begin{tabular}{@{}c@{}} $\Delta K$ [mag] \\ \end{tabular} \\

\hline

HD~128220 & $527^{+16}_{-16}$ & (+0.97, --2.14) & (0.17, 0.01) & --23.2 & 1.65 & $1.24\pm0.09$ & $2.9\pm0.4$ & $3.84\pm0.15$ \\

BD\,$-07^\circ5977$* & $754^{+15}_{-12}$ & (+1.79, --3.30) & (0.32, 0.11) & --30.9 & 2.94 & $2.83\pm0.22$ & $1.9\pm0.2$ & $4.30\pm0.11$ \\

BD\,$+10^\circ2357$ & $170.0^{+1.0}_{-1.1}$ & (--5.88, +2.87) & (0.06, 0.03) & --36.8 & 8.44 & $1.11\pm0.01$ & $7.5\pm0.2$ & $2.81\pm0.03$ \\

HD 283048 & $421^{+10}_{-11}$ & (+4.75, +4.83) & (0.02, 0.01) & --11.3 & 12.25 & $2.85\pm0.07$ & $9.8\pm0.1$ & $2.52\pm0.01$ \\

TYC 1703-394-1 & $350.4^{+3.5}_{-3.1}$ & (--6.34, +2.42) & (0.03, 0.02) & +9.9 & 1.98 & $2.38\pm0.03$ & $4.7\pm 0.1$ & $3.32\pm0.02$ \\

CPD\,$-71^\circ172$ & $326.5^{+1.8}_{-1.8}$ & (+2.36, --4.30) & (0.03, 0.01) & --32.9 & 3.42 & $1.60\pm0.01$ & $5.2\pm0.1$ & $3.21\pm0.02$ \\

\hline

\end{tabular*}
\tablefoot{For every target, we list: the distance ($d$) from \citet{2021BailerJones}; the best-fit position of the detected companion ($\Delta E,\,\Delta N$); the major axis ($e_{\rm maj}$), minor axis ($e_{\rm min}$), and position angle of the major axis east of north (PA) for the position error ellipse; the reduced chi-squared statistic ($\chi^2_{\rm red}$) ; the instantaneous projected separation ($\rho$); the $K$-band flux ratio of the companion relative to the MS component ($f_2/f_1$) expressed in percent; and the corresponding magnitude contrast $\Delta K = K_{\rm sdO/B} - K_{\rm MS}$. *See additional notes on BD\,$-07^\circ5977$ in Appendix\,\ref{app:notes}.}
   \end{table*}

\begin{figure*}
    \centering
    \includegraphics[width=0.85\linewidth]{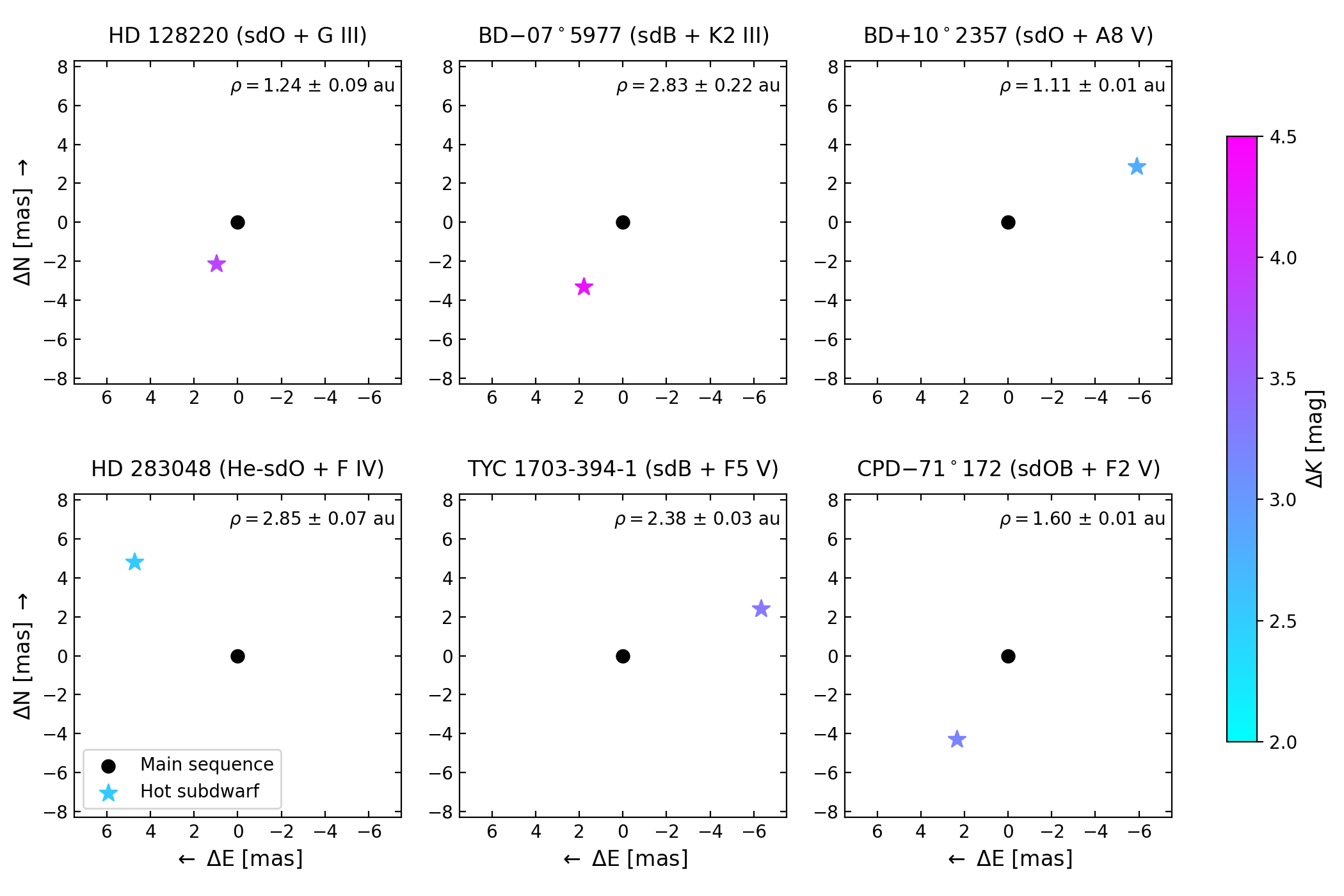}
    \caption{Relative astrometric positions of composite sdO/B + MS binaries resolved with GRAVITY. Each panel corresponds to one target, showing (i) the MS component (black circle) fixed at origin, and (ii) the sdO/B component (colored star) at its best-fit position in the sky plane. The color of the sdO/B indicates the magnitude contrast $\Delta K = K_{\rm sdO/B} - K_{\rm MS}$ (see Table\,\ref{tab:results}). Error ellipses are smaller than the sdO/B marker size. Also, for each panel, shown in the title are component spectral types (see Appendix\,\ref{app:notes}), and in the top right are the instantaneous physical projected separations.}
    \label{fig:all}
\end{figure*}

Given the composite systems in our sample are pairs of hot sdO/Bs and cool MS stars, the latter dominate in the $K$ band. Consequently, in each system, we attempt to search for the faint companion, i.e. the sdO/B. We adopt the CANDID algorithm \citep{2015Gallenne} implemented in the PMOIRED software \citep{2022Merand} for analysis of interferometric data. 
Each observation consists of multiple exposures, leading to multiple |V|-T3PHI datasets per target. The change in telescope baselines due to the earth's rotation or change in the relative positions of the binary components can cause |V|-T3PHI to change over time. Given the typical periods of $\approx10^3$\,d for sdO/B + MS binaries, we assume the binary positions do not change, and model all exposures for a target simultaneously taking the earth's rotation into account.
To search for the companion, we first fix the bright star (presumably the MS component) at the origin, and fix its flux to 1 ($f_1=1$). We then define a grid on the sky-plane covering the ranges $\Delta E = -100\,\,{\rm to}\,\,100$ mas and $\Delta N = -100\,\,{\rm to}\,\,100$ mas in steps of 1 mas each to span the entire GRAVITY sensitivity region and search for the companion. Both components are modeled as unresolved uniform disks with an angular diameter of 0.2 mas, well below the resolution limit of GRAVITY. This assumption is justified given all targets are a few hundred parsecs away (see Table\,\ref{tab:results}), resulting in angular sizes of the order of 0.01 mas for low mass MS stars and even smaller for sdO/Bs. Finally, although the two components are expected to have different spectral slopes due to different temperatures, we adopt a flat spectrum for both, which is a reasonable assumption to constrain binary/multiple systems \citep[e.g.,][]{2024Deshmukh}.

For the grid search, the companion position is allowed to vary, along with its flux ($f_2$). The best-fit solution is determined by minimizing the reduced chi-squared statistic for the |V|-T3PHI data. We also require an $8\sigma$ detection level (with a single star model as the reference fit) to avoid numerical degeneracies and confidently claim the detection of a spatially resolved companion. In each of our six targets, we successfully detect the sdO/B companion to the MS component.
 
We obtain the relative position and corresponding error ellipse of the sdO/B companion, and the companion flux ratio $f_2/f_1=f_2$. We also compute projected physical separation based on the measured angular separation and distance to the target \citep[geometric distance from][]{2021BailerJones}, assuming Gaussian $1\sigma$ errors. A summary of the best-fit solutions is given in Table\,\ref{tab:results} (all errors are $1\sigma$). Figure\,\ref{fig:all} shows the relative positions of the binary components for all targets. A collection of the best model fits for the GRAVITY data along with individual target notes is provided in Appendix\,\ref{app:notes}. We also explored three-component models to search for potential tertiary companions, finding a preliminary tertiary candidate in BD\,$+10^\circ2357$ which can only be confirmed with follow-up observations. More details on tertiary search are provided in Appendix\,\ref{app:triple}.

Additionally, we performed an SED analysis using archival photometric measurements of each system to obtain an independent estimate of the $K$-band magnitudes for both components. This is briefly described in Appendix\,\ref{app:sed}, and the comparison with GRAVITY results is listed in Table\,\ref{tab:sed}.

\section{Discussion and conclusion}
\label{sec:dis}

We present the first interferometric observations of six wide sdO/B binaries with VLTI/GRAVITY, all of which were spatially resolved into two components. We also find a candidate tertiary companion in BD\,$+10^\circ2357$, pending confirmation (Appendix\,\ref{app:triple}). This campaign successfully demonstrates the efficiency and precision of long-baseline interferometry in resolving wide sdO/B binaries with magnitude contrasts up to $\Delta K\approx5$. The instantaneous projected physical separations for all six binaries range from $1-3$ au, broadly consistent with the orbital major axis estimates for wide sdO/B binaries reported by \citet{2017Vos}. Subsequently, all systems could have resulted from the stable mass transfer channel of sdO/B formation. BD\,$-07^\circ5977$ is consistent with this channel \citep{2017Vos}, while HD~128220 is likely a post-asymptotic giant branch binary \citep{1993Rauch}. Although these two systems have known spectroscopic orbital solutions, at least one more interferometric observation or an astrometric orbit from {\it Gaia} is necessary to determine their 3D orbits and dynamical masses. For the remaining four systems, ongoing spectroscopic monitoring is expected to yield spectroscopic orbital solutions in the near future, that can be combined similarly with {\it Gaia} and multi-epoch interferometric observations. Table\,\ref{tab:stat} summarizes the current status and necessary observations to determine 3D orbits for all systems.

Epoch astrometry from {\it Gaia} DR4 is predicted to detect the photocenter motion in many sdO/B + MS binaries \citep{2026Nagarajan}, which by itself will not be sufficient to constrain their true orbits.
Using SED analysis, flux contribution of each component in different bands can be estimated, enabling conversion of {\it Gaia} photocenter motion into a relative astrometric orbit. However, systematic uncertainties in the SED models propagate into the inferred masses. For example, the sdO/B $K$-band magnitudes from SED fits are broadly consistent with, but differ from, those derived from the GRAVITY analysis (Table\,\ref{tab:sed}).

Direct measurements of the relative binary positions with interferometry evade the need for SED models, and instead directly set a precise physical scale for the binary orbits. Furthermore, for binaries resolved with interferometry but not detected with {\it Gaia} astrometry, multi-epoch interferometric observations will be crucial for 3D orbit determination.

Measuring accurate model-independent masses of both components will be pivotal in understanding the formation of sdO/Bs in such binaries, and in calibrating other methods of mass determination. With a growing sample size \citep[e.g.][]{2026Molina}, one can also expect increasingly robust emergence of trends and correlations between parameters such as orbital period, eccentricity, rotational velocity, mass ratio, etc. that are intimately connected to stable mass transfer physics \citep{2026Parkosidis}. By providing a means to measure model-independent masses, long-baseline interferometry will thus play an important and unique role in studying the formation of wide sdO/B binaries.

   \begin{table}[ht]
      \caption[]{Status of orbital constraints for our sdO/B + MS sample.}
         \label{tab:stat}
         \renewcommand{\arraystretch}{1.15}
\begin{tabular*}{\columnwidth}{@{\extracolsep{\fill}} cccc }
\hline
\begin{tabular}{@{}c@{}} Target  \end{tabular} 
& \begin{tabular}{@{}c@{}} Known orbital params. \end{tabular} 
& \begin{tabular}{@{}c@{}} Data needed \\ \end{tabular} \\

\hline

HD~128220 & $P,e,\omega,T_0,a\,{\rm sin}\,i, q$ $^{(a)}$ & I/A  \\

BD\,$-07^\circ5977$ & $P,e,\omega,T_0,a\,{\rm sin}\,i, q$ $^{(b)}$ & I/A  \\

BD\,$+10^\circ2357$ & none & S + I/A \\

HD~283048 & none & S + I/A \\

TYC~1703-394-1 & none & S + I/A  \\

CPD\,$-71^\circ172$ & none & S + I/A  \\

\hline

\end{tabular*}
\tablefoot{Listed are currently known orbital parameters and data needed to derive 3D orbits for all targets; observing techniques include interferometry (I), astrometry (A) and spectroscopy (S). $^{(a)}$\citet{1990Howarth}, $^{(b)}$\citet{2017Vos}. }
   \end{table}

\bibliographystyle{aa} % style aa.bst
\bibliography{aanda} % your references WR.bib

\begin{appendix}

\counterwithin{figure}{section}

\section{Acknowledgments}

\begin{acknowledgements}
We thank the anonymous referee for their comments and suggestions that helped improve this work. This research is based on observations collected at the European Southern Observatory under ESO program ID 116.28ZD (PI. Deshmukh). This research received support from Long term structural funding - Methusalem funding by the Flemish Government (project SOUL, METH/24/012) at KU Leuven. HD was supported by the Deutsche Forschungsgemeinschaft (DFG) through grants GE2506/17-1 and GE2506/9-2. MD was supported by the Deutsches Zentrum für Luft- und Raumfahrt (DLR) through grant 50-OR-2510. AP acknowledges support from the Research Foundation - Flanders (FWO), grant agreement No. 11M8325N (PhD Fellowship). This research was supported by Deutsche Forschungsgemeinschaft (DFG, German Research Foundation) under Germany’s Excellence Strategy - EXC 2121 "Quantum Universe" – 390833306. Co-funded by the European Union (ERC, CompactBINARIES, 101078773). Views and opinions expressed are however those of the author(s) only and do not necessarily reflect those of the European Union or the European Research Council. Neither the European Union nor the granting authority can be held responsible for them. MV acknowledges support from the Fondecyt Regular grant No. 1211941 and 1250525.

\end{acknowledgements}

\section{Summary of observations}
\label{app:obs}

   \begin{table}[ht]
      \caption[]{GRAVITY observations of our sdO/B + MS sample.}
         \label{tab:obs}
         \renewcommand{\arraystretch}{1.15}
\begin{tabular*}{\columnwidth}{@{\extracolsep{\fill}} cccc }
\hline
\begin{tabular}{@{}c@{}} Target  \end{tabular} 
& \begin{tabular}{@{}c@{}} Calibrator \end{tabular} 
& \begin{tabular}{@{}c@{}} Array \end{tabular}
& \begin{tabular}{@{}c@{}} BJD -- 2.4e6 \\ \end{tabular} \\

\hline

HD~128220 & TYC~1481-559-1 & AT & 61162.264 \\

BD\,$-07^\circ5977$ & HD~220502 & AT & 60977.604 \\

BD\,$+10^\circ2357$ & BD\,$+01^\circ2613$ & AT & 61090.847 \\

HD~283048 & BSD\,47-534 & UT & 60990.688 \\

TYC~1703-394-1 & TYC~1109-1354-1 & UT & 60952.543 \\

CPD\,$-71^\circ172$ & TYC~9142-736-1 & UT & 60953.731 \\

\hline

\end{tabular*}
\tablefoot{We list for each target the calibrator star, the VLTI array used and the Barycentric Julian Date (BJD) of start of observation.}
   \end{table}

\section{Notes on individual targets}
\label{app:notes}

Here we discuss individual targets briefly and provide the best-fit plots for our GRAVITY observations.

\begin{figure}
    \centering
    \includegraphics[trim={24cm 1cm 0cm 1cm}, clip,width=0.95\linewidth]{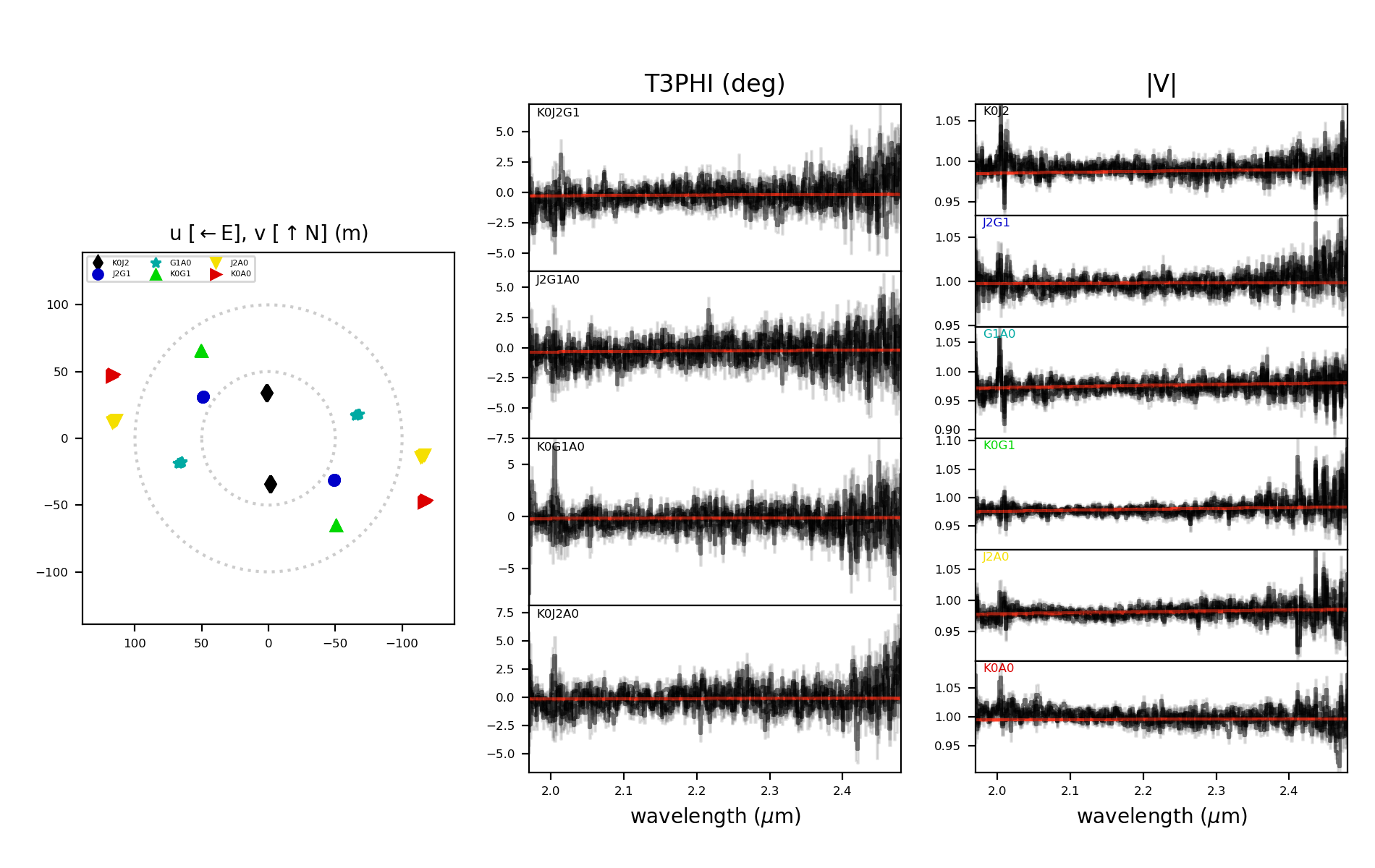}
    \caption{GRAVITY data (black) along with the best-fit model (red) for HD~128220. The $1\sigma$ error regions are shown in gray (seen more clearly in Fig.\,\ref{fig:bd-075977}). The left panel shows the closure phase (T3PHI) in degrees, while the right panel shows the visibility amplitude (|V|), both as function of wavelength. Each sub-panel also shows the closure triangle/baseline for T3PHI/|V| in the top left corner. The sub-panels for |V| are arranged from shortest to longest baseline top to bottom.}
    \label{fig:hd128220}
\end{figure}

\textbf{HD~128220} is an sdO + GIII binary that has an orbital solution first reported by \citet{1990Howarth}. The orbital period is $872\pm1$ days, while the eccentricity is $0.21\pm0.01$, consistent with other wide sdO/B binaries. However, the sdO in HD~128220 is likely a post-asymptotic giant branch star based on detailed ultraviolet spectroscopic analysis by \citet{1993Rauch}. In our analysis of GRAVITY data, we successfully resolved the two components. The secondary flux contribution is relatively low within the sample, leading to higher uncertainties. Figure\,\ref{fig:hd128220} shows the best-fit model to the GRAVITY data along with corresponding residuals.

\begin{figure}
    \centering
    \includegraphics[trim={8.5cm 0.46cm 0cm 1cm}, clip, scale=0.56]{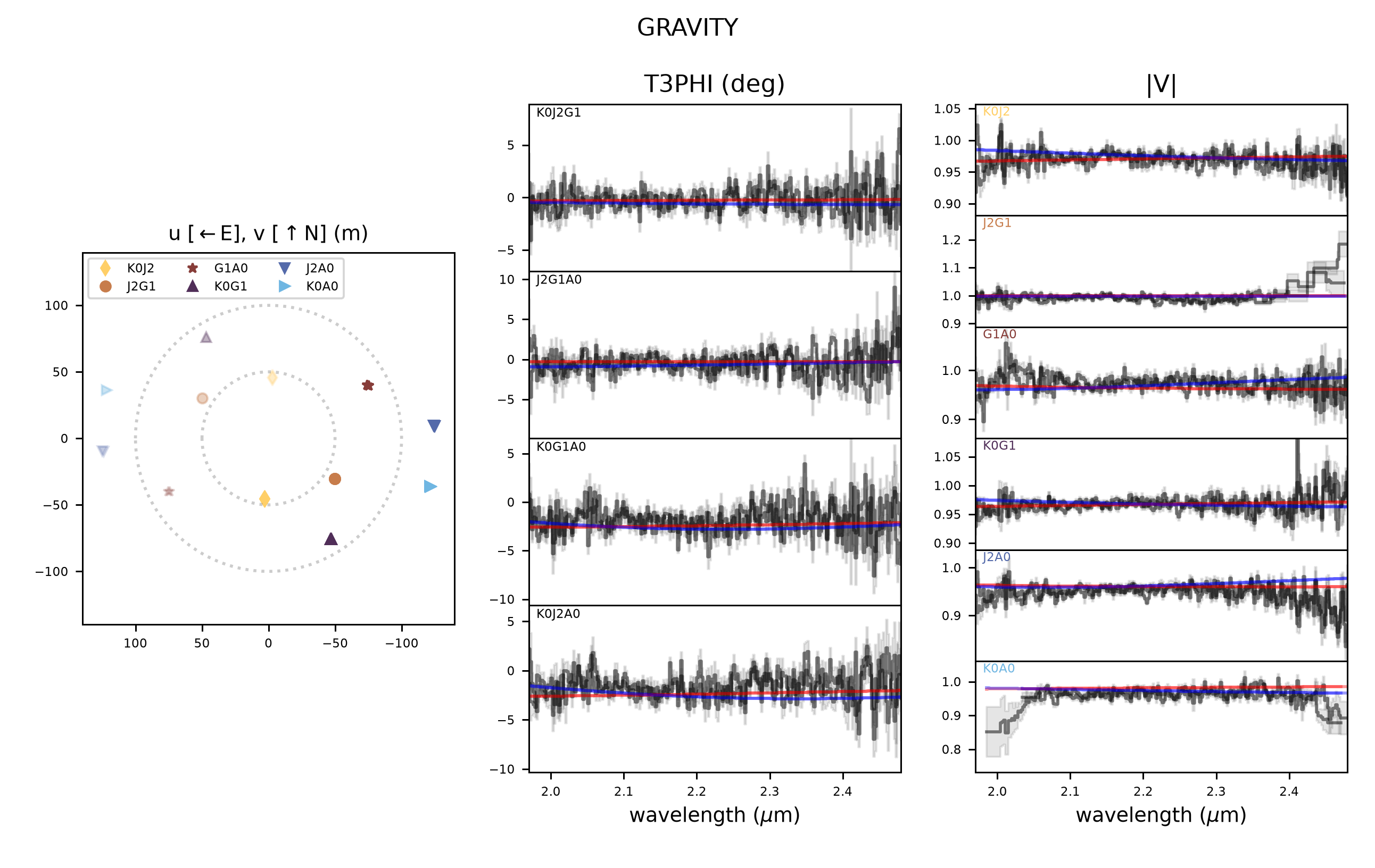}
    \caption{GRAVITY data for BD\,--07$^\circ$5977, similar to Fig.\,\ref{fig:hd128220}. Both solutions discussed in Appendix\,\ref{app:notes} are shown; the $(\Delta E, \Delta N)$ = (1.79, --3.30) mas solution in red, and the $(\Delta E, \Delta N)$ = (--4.65, 6.70) mas solution in blue. The former is reported in Table\,\ref{tab:results}.}
    \label{fig:bd-075977}
\end{figure}

\textbf{BD\,--07$^\circ$5977} is an sdB + K2III binary with a known spectroscopic orbital solution \citep{2017Vos}. It has an orbital period of $1262\pm1$ days and an eccentricity of $0.16\pm0.01$. The flux contribution of the sdB is the lowest in the sample (Table\,\ref{tab:results}), leading to a marginal (but significant) detection in GRAVITY data (see Fig.\,\ref{fig:bd-075977}). 

We find two possible solutions for the relative position of the sdB: $(\Delta E, \Delta N)$ = (1.79, --3.30) and (--4.65, 6.70) mas, both with near-identical sdB flux contribution. However, the second solution corresponds to a projected physical separation $\rho\approx6.1\,{\rm au}$, which is almost twice the maximum binary separation $a\,(1+e)\approx3.1\,{\rm au}$ estimated by \citet{2017Vos}. We therefore select the first solution, (1.79, --3.30) mas, and report it in Table\,\ref{tab:results}.

\begin{figure}
    \centering
    \includegraphics[trim={8.5cm 0.46cm 0cm 1.07cm}, clip, scale=0.56]{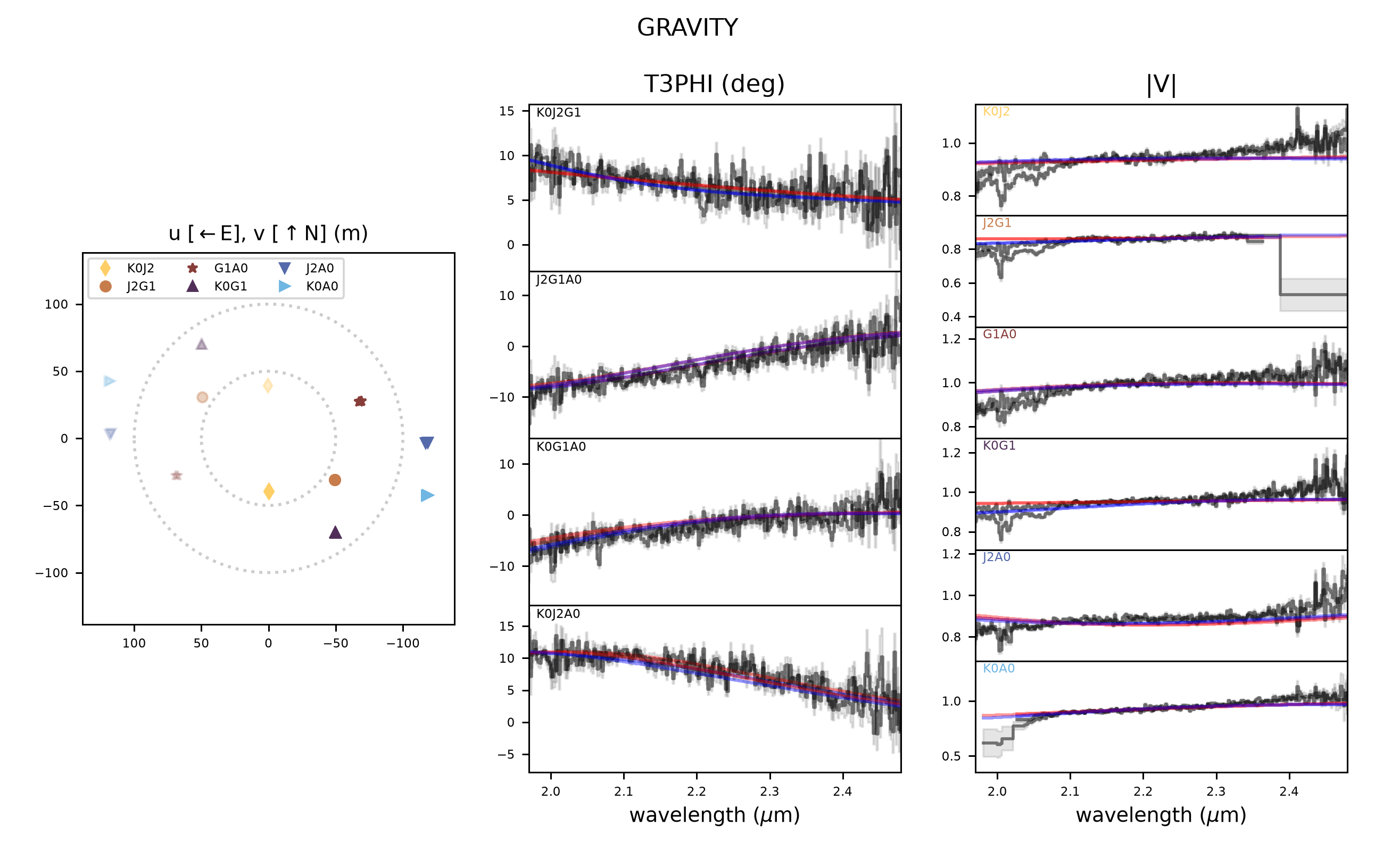}
    \caption{GRAVITY data for BD\,+10$^\circ$2357, similar to Fig.\,\ref{fig:hd128220}. The best-fit two-component model is shown in red, while the three-component model discussed in Appendix\,\ref{app:triple} is shown in blue.}
    \label{fig:bd+102357}
\end{figure}

\textbf{BD\,+10$^\circ$2357} was initially identified as an sdO + A binary by \citet{1980Berger}. It was updated to sdO + A8 V using SED analysis as part of the 500-pc volume-limited sample compiled by \citet{2024Dawson}, although the system does not have an orbital solution yet. Furthermore, \citet{2022Krzesinski} note the presence of pulsations in TESS observations of BD\,+10$^\circ$2357 and speculate their likely origin to be the A8 V component. The binary was robustly resolved in our GRAVITY observations, as shown by the best-fit model in Figure\,\ref{fig:bd+102357}. It has a relatively high $\chi^2_{\rm red}$ value (Table\,\ref{tab:results}), primarily due to relatively poor |V| calibration (|V| slightly exceeding 1) around 2.4 $\mu$m and noisy |V| data around 2.0 $\mu$m. We still use the full wavelength range in our fits given the T3PHI data are not affected. For a sanity check, we also repeated our companion search only using the 2.1--2.35 $\mu$m range and obtained near-identical fit parameters.

A tentative tertiary component was also detected, discussed in Appendix\,\ref{app:triple} and also shown in Figure\,\ref{fig:bd+102357}.

\begin{figure}
    \centering
    \includegraphics[trim={24cm 1cm 0cm 2.5cm}, clip,width=0.95\linewidth]{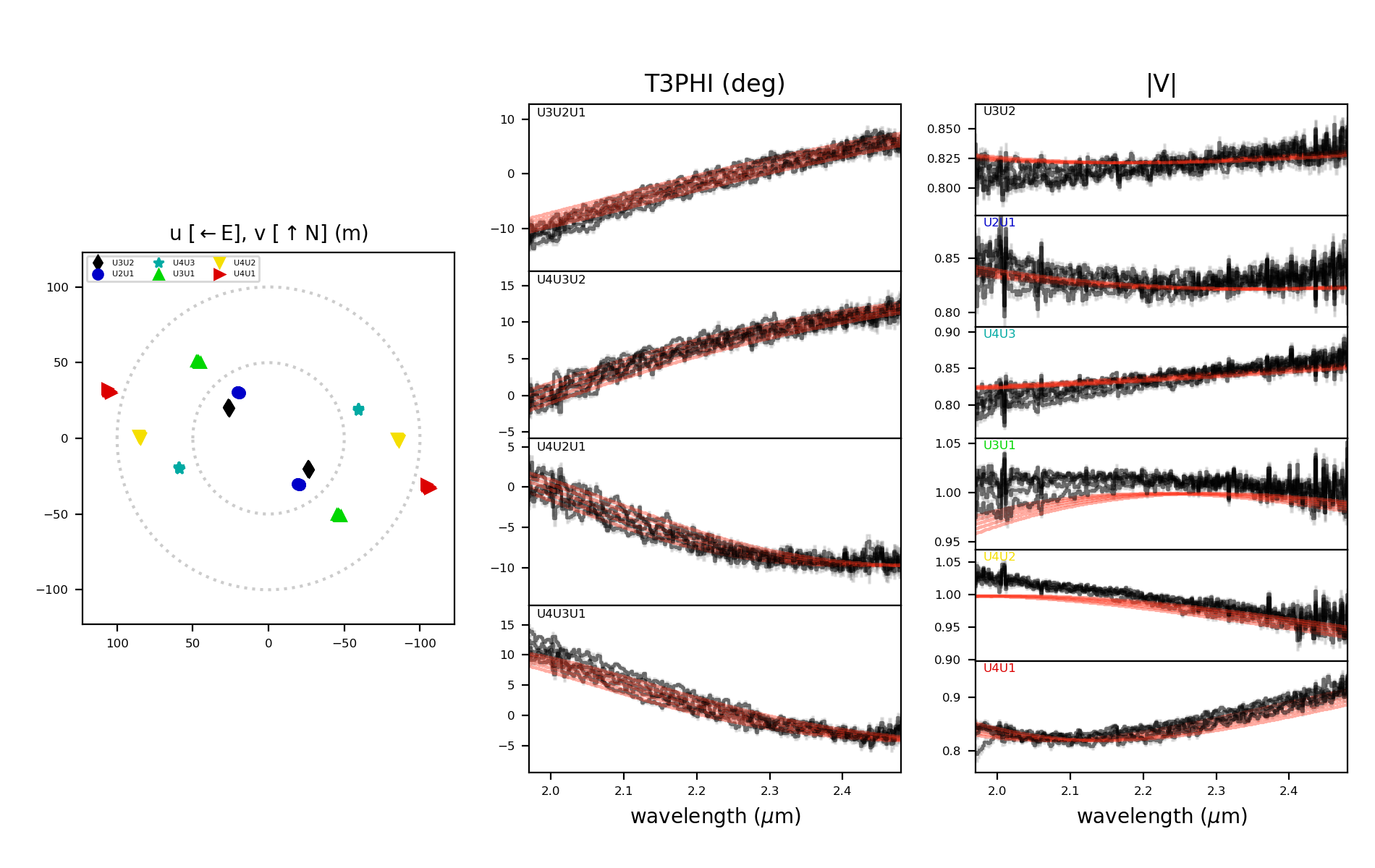}
    \caption{GRAVITY data for HD 283048, similar to Fig.\,\ref{fig:hd128220}. Multiple datasets corresponding to multiple exposures, as discussed in Sect.\,\ref{sec:comp}, are most apparent here. The best-fit model also shows multiple red lines accordingly.}
    \label{fig:hd283048}
\end{figure}

\textbf{HD 283048} was first studied several decades ago by \citet{1978Laget}, who classified the system as a potential hot subluminous star with a late F-type companion. More recently, it was classified as a He-sdO + F IV binary based on SED analysis by \citet{2024Dawson}. We successfully resolve the binary with GRAVITY, as shown in Figure\,\ref{fig:hd283048}. Similar to BD\,+10$^\circ$2357, it also has a high $\chi^2_{\rm red}$ value (Table\,\ref{tab:results}), also due to poor |V| calibration (|V|>1) between 2.0--2.2 $\mu$m in the U3U1 and U4U2 baselines.
An orbital solution is still pending for the system.

\begin{figure}
    \centering
    \includegraphics[trim={24cm 1cm 0cm 2.5cm}, clip,width=0.95\linewidth]{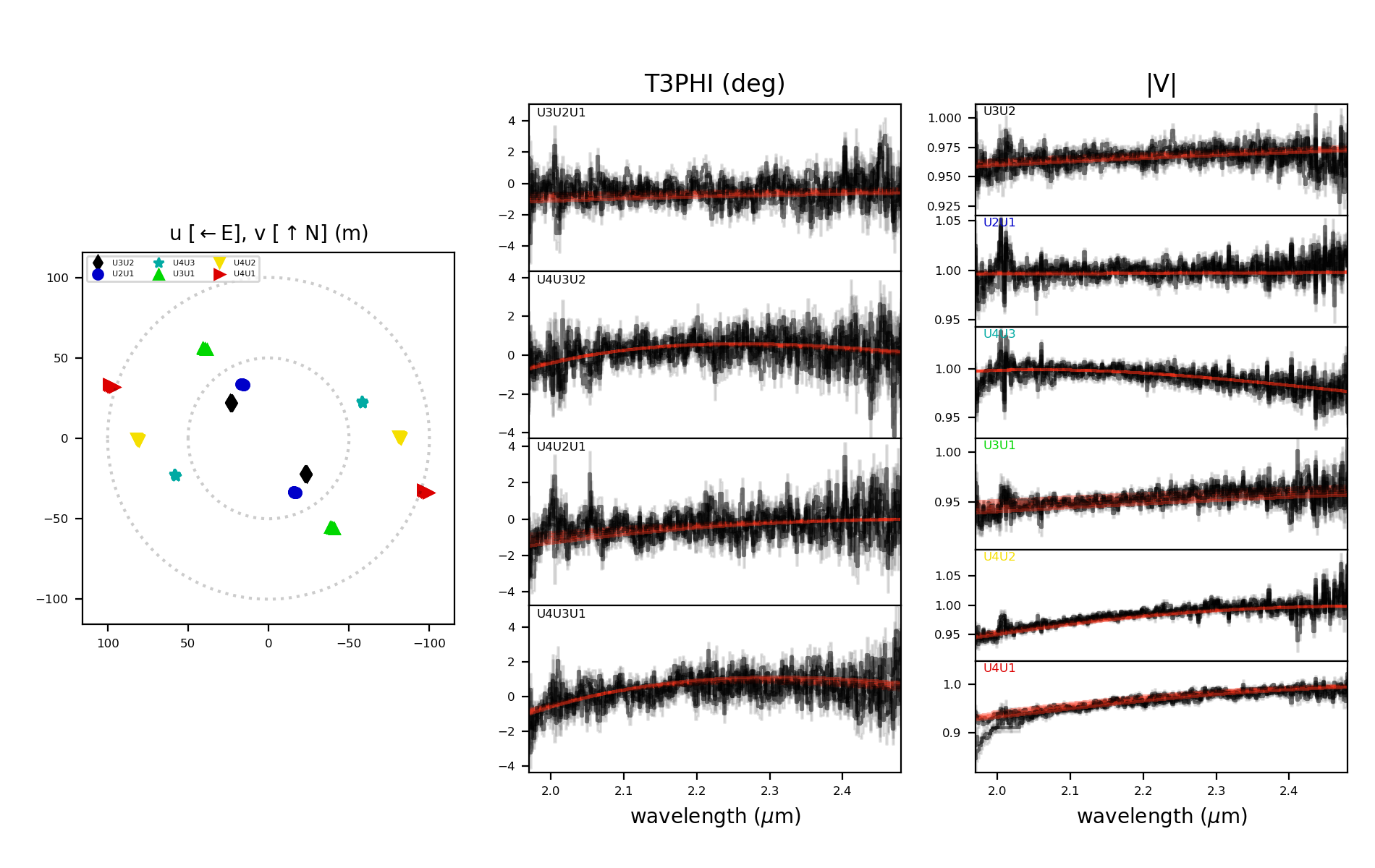}
    \caption{GRAVITY data for TYC1703-394-1, similar to Fig.\,\ref{fig:hd128220}.}
    \label{fig:tyc}
\end{figure}

\textbf{TYC 1703-394-1} (or GALEX J222758.5+200623) was first identified as an sdO/B with the {\it GALEX} ultraviolet mission. It is now classified as an sdB + F5 V composite binary \citep{2024Dawson}, although an orbital solution is yet to be determined. We resolve the binary with GRAVITY, with the detection being marginal but significant (see Fig.\,\ref{fig:tyc}).

\begin{figure}
    \centering
    \includegraphics[trim={24cm 1cm 0cm 1cm}, clip,width=0.95\linewidth]{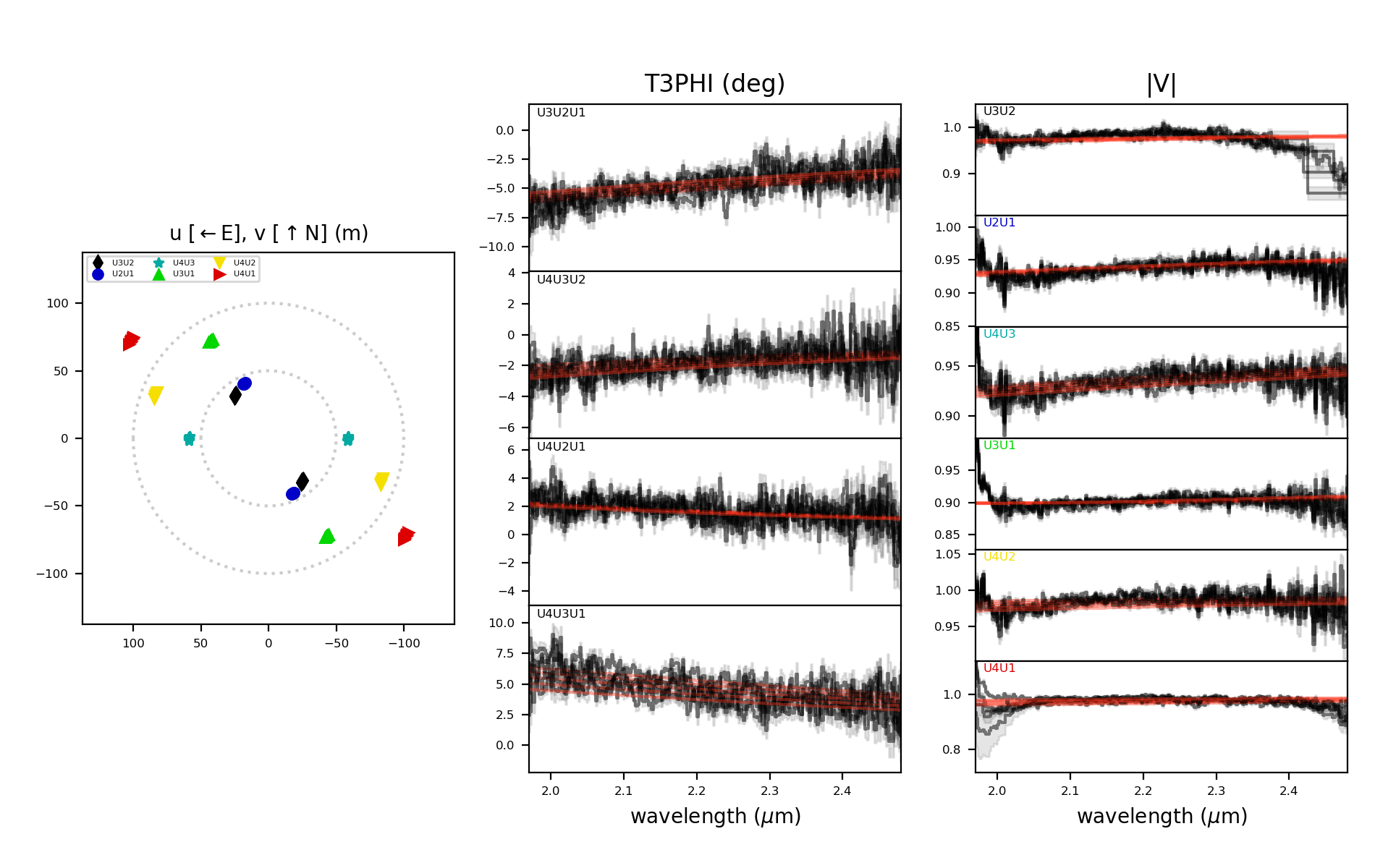}
    \caption{GRAVITY data for CPD\,$-71^\circ172$, similar to Fig.\,\ref{fig:hd128220}.}
    \label{fig:cpd-71172}
\end{figure}

\textbf{CPD\,--71$^\circ$172} was classified as an sdOB + F-type binary based on UV-optical observations by \citet{1987Viton}. It was recently updated to sdOB + F2 V by \citet{2024Dawson}, with an orbital solution still not known. The binary is clearly resolved with GRAVITY, as shown in Figure\,\ref{fig:cpd-71172}.

\section{Tertiary companion search}
\label{app:triple}

We also explored the possibility of a third component being present in our targets. The same grid search approach as described in Section\,\ref{sec:comp} was used, this time with a third unresolved uniform disk added to the model. We adopted the best-fit binary solution for the MS and sdO/B components as the initial guess, while a grid search was performed for the third component over the ranges $\Delta E = -100\,\,{\rm to}\,\,100$ mas and $\Delta N = -100\,\,{\rm to}\,\,100$ mas in steps of 1 mas each, with its flux contribution allowed to vary. 

No significant tertiary component was detected in five targets. The sixth target, BD\,+10$^\circ$2357, showed a significant improvement in the chi-squared statistic accompanied by a 7$\sigma$ tertiary component detection, just falling short of our $8\sigma$ detection threshold. The third component is detected at ($\Delta E, \Delta N$) = (3.93, 11.21) mas, with a relative flux of $3\pm1\%$ with respect to the MS component.  
We do not confirm BD\,+10$^\circ$2357 as a triple system yet based on these results, but establish it as a strong candidate. Future interferometric observations of the system and astrometric data from {\it Gaia} DR4 might enable us to robustly identify and characterize this potential third component.

\section{Spectral energy distributions}
\label{app:sed}

\begin{figure*}%[H]
    \centering
    \includegraphics[width=0.48\textwidth]{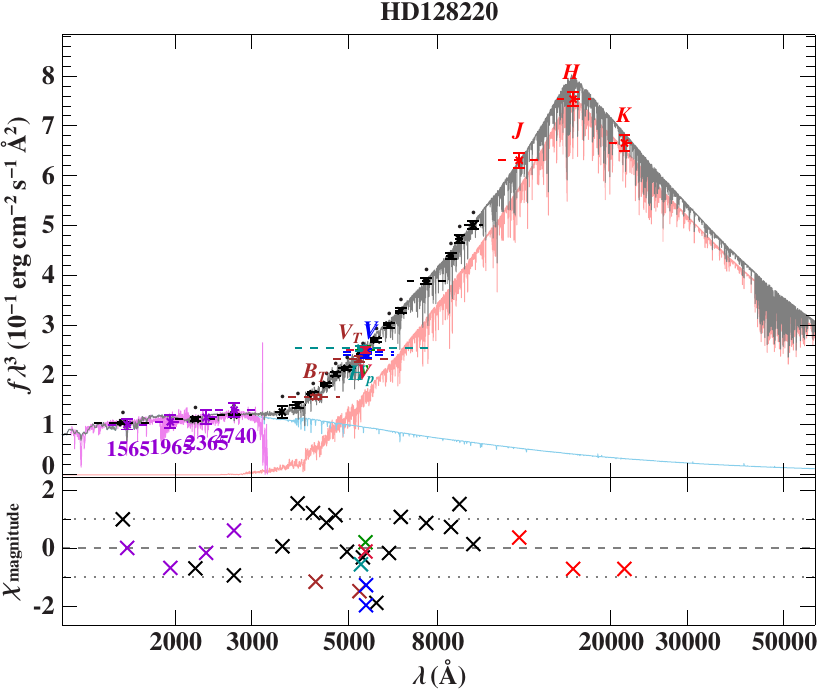}
    \includegraphics[width=0.48\textwidth]{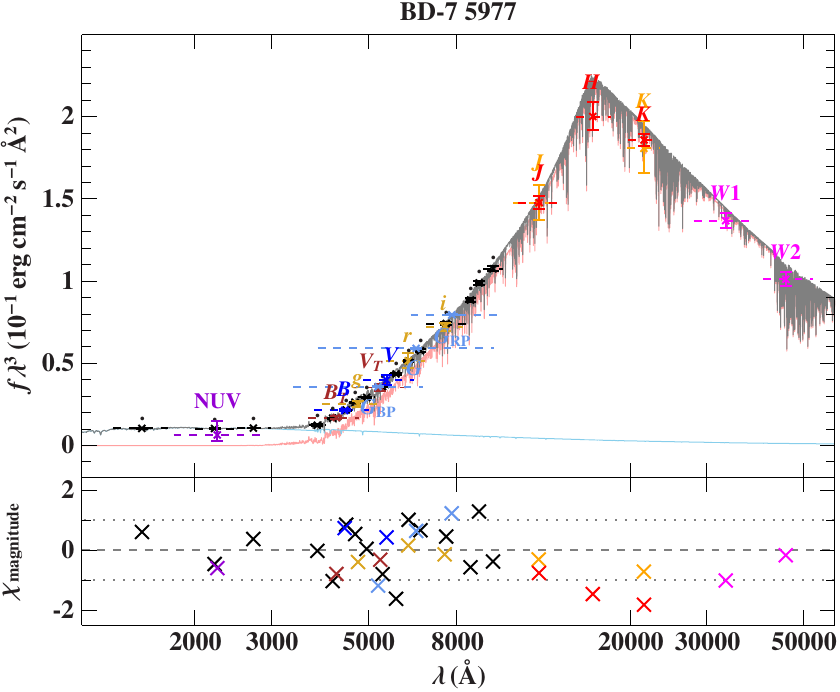}\\[3pt]
    \includegraphics[width=0.48\textwidth]{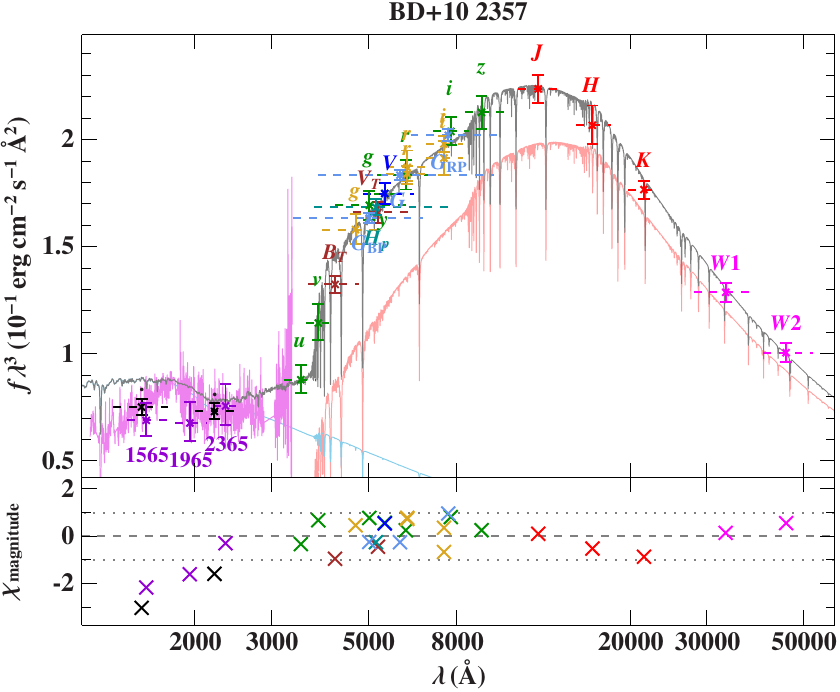}
    \includegraphics[width=0.48\textwidth]{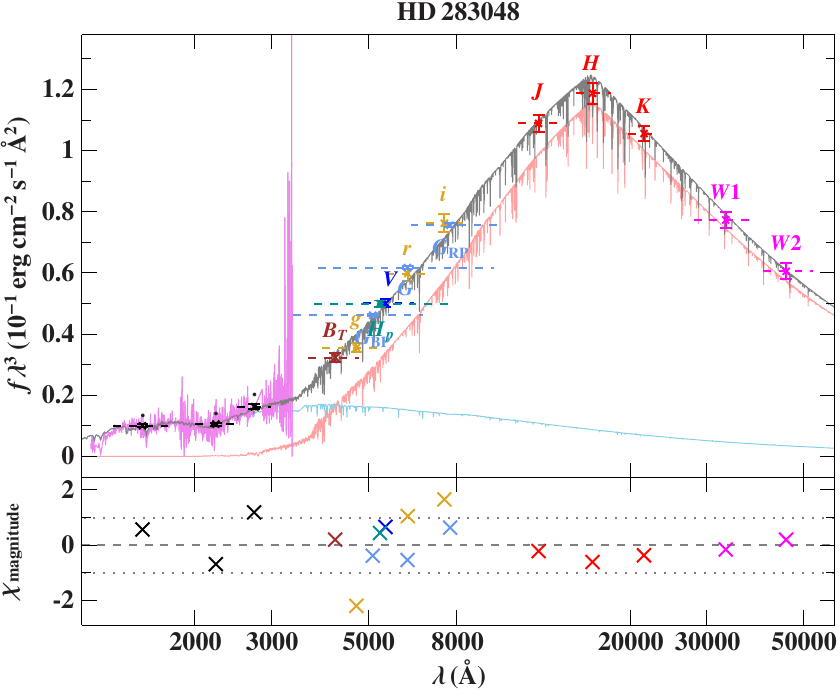}\\[3pt]
    \includegraphics[width=0.48\textwidth]{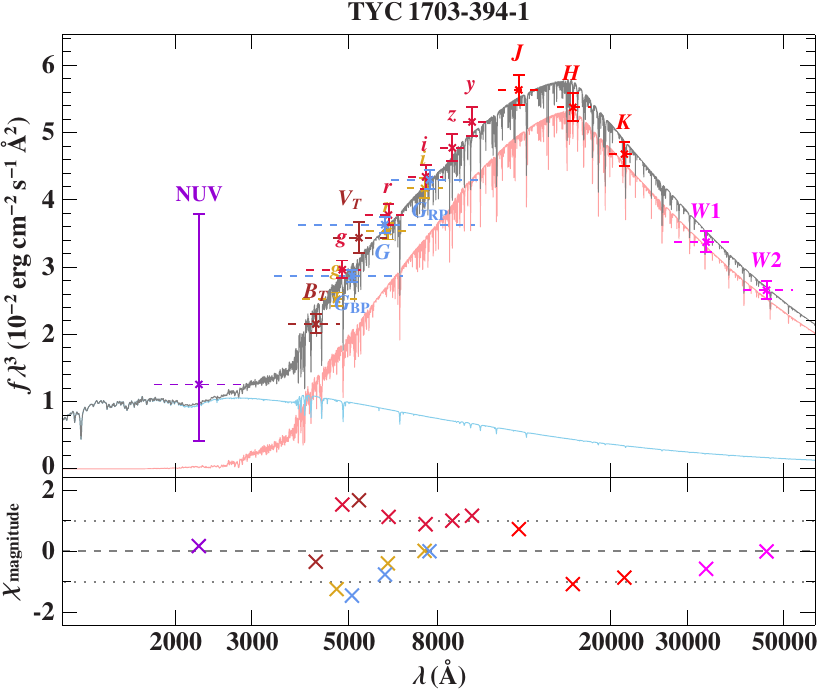}
    \includegraphics[width=0.48\textwidth]{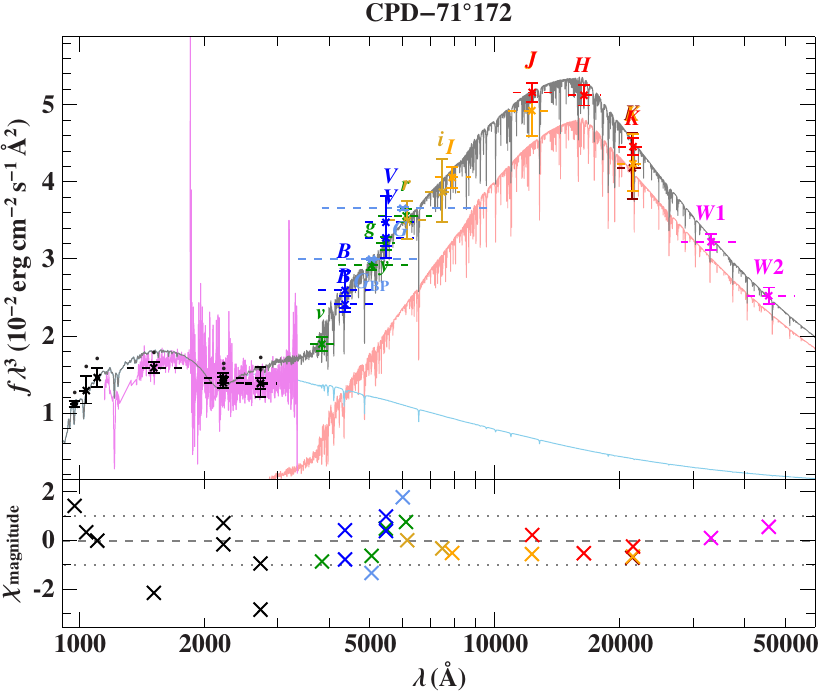}%\\
    \caption{Composite SED fits; the sdO/B component is blue, the cool companion is red, and the combined model is grey. 
    Where available, {\it IUE} low-resolution spectra are shown in pink \citep{IUE_INES_1999}, while photometric measurements are marked by colour: \textit{Gaia} XP (black), \textit{Gaia} BP/RP \citep[light blue]{GaiaDR3_SED}, {\it GALEX} \citep[purple]{SED_galex_ais}, Johnson \cite[blue]{APASS_2015}, PanSTARRS \citep[dark red]{PS1_DR2_2020}, SkyMapper \citep[green]{Skymapper2024}, 2MASS \citep[red]{Skrutskie2006_2MASS}, {\it WISE} \citep[pink]{2019Schlafly}, Tycho \citep[brown]{Tycho2_2000}, DENIS \citep[orange]{DENIS_2005}. The bottom panels show uncertainty-scaled residuals. 
    }
    \label{fig:sed}
\end{figure*}

We performed spectral energy distribution (SED) fits to (unresolved) photometric measurements of each system, following the method of \cite{2018Heber}; the SEDs are shown Fig.\ \ref{fig:sed}. 
This allows an independent estimate of the $K$-band magnitudes for both the MS and sdO/B components. 
GRAVITY magnitudes are based on low-resolution spectra in the $K$-band, rather than a well-defined photometric filter passband. 
To facilitate an approximate comparison, we integrated the best-fit SED models for both stars in each system over the GRAVITY spectral range (1.97 -- 2.45\,$\mu$m). 
These magnitudes are compared to GRAVITY flux ratio measurements, which were converted to magnitude differences and are anchored to the same total flux as the SED model. The results are stated in Table \ref{tab:sed}. As expected, the agreement is good for the MS component, and worse for the sdO/B component. This is likely because the SED-based $K$-band magnitudes for the sdO/B components are rather uncertain, given that the sdO/Bs contribute little even in the optical range.

\begin{table}[H]
\caption{
Comparison of $K$-band magnitudes obtained from SED fits and GRAVITY. 
}
\label{tab:sed}
\centering
\setstretch{1.1}
\resizebox{\columnwidth}{!}{
\begin{tabular}{lrrrr}
\toprule
\toprule
System & $K^\mathrm{SED}_\mathrm{MS}$ & $K^\mathrm{SED}_\mathrm{sdO/B}$ & $K^\mathrm{G}_\mathrm{MS}$ & $K^\mathrm{G}_\mathrm{sdO/B}$\\
 &  [mag] &  [mag] & [mag]  & [mag]  \\
\midrule
HD~128220 & $8.9\pm 0.1$ & $12.2\pm 0.1$ & $8.9\pm0.1$ & $12.7\pm0.2$ \\
BD\,$-07^\circ5977$ &  $10.2\pm 0.1$ & $14.9\pm 0.1$ & $10.2\pm0.1$ & $14.5\pm0.2$ \\
BD\,$+10^\circ2357$ &  $10.4\pm 0.1$ & $13.3\pm 0.2$ & $10.4\pm0.1$ & $13.2\pm0.1$ \\
HD 283048 &  $10.9\pm 0.1$ & $13.8\pm 0.1$ & $10.9\pm0.1$ & $13.5\pm0.1$ \\
%tot:   10.865 -0.275 +0.274
%c1:    13.829 -0.192 +0.202
%c2:    10.938 -0.284 +0.284
%c1-c2: 2.891 -0.200 +0.210
TYC 1703-394-1 &  $11.8\pm 0.1$ & $14.6\pm 0.2$ & $11.8\pm0.1$ & $15.1\pm0.1$ \\
CPD\,$-71^\circ172$ & $11.9\pm0.1$ & $14.4\pm 0.3$ & $11.9\pm0.1$ & $15.1\pm0.1$ \\
\bottomrule
\end{tabular}
}
\tablefoot{
Uncertainties are stated as statistical only for GRAVITY.
}
\end{table}

\clearpage

\end{appendix}

\end{document}